 \documentclass[preprint, 12pt]{elsarticle}

 \usepackage{amssymb}

 \usepackage{amsmath}

 \newtheorem{theorem}{Theorem}

 \newtheorem{defi}{Definition}
 
 \newtheorem{property}{Property}
  \newtheorem{procedure}{Procedure}
  
 \journal{Journal Name}

\begin{document}
	
	 \begin{frontmatter}
		
		

	 \title{Definition and Analytical Expression on State Observe Ability for Linear Discrete-time Systems with the Bounded Noise Energy
		 \thanks{Work supported by the National Natural Science Foundation of China (Grant No. 61273005)}}
		
		
		 \author{Mingwang Zhao}
		
		 \address{Information Science and Engineering School, Wuhan University of Science and Technology, Wuhan, Hubei, 430081, China \\
			Tel.: +86-27-68863897 \\
		Work supported by the National Natural Science Foundation of China (Grant No. 61273005)}
		
		 \begin{abstract}
			In this article, the definition on the observe ability and its relation to the signal detecting  performance  are studied systematically for the linear discrete-time(LDT) systems. Firstly, to define and  analyze the observe ability for the practical systems with the  measured noise, six kinds of bounded noise models are classified. For the noise energy bounded case, the  observability ellipsoid and the image 
			observability ellipsoid are defined by the state observed error and then a novel concept on the LDT systems, called as the observe ability, is proposed. Based on that, some theorems and properties about the observe ability and the signal detecting performances  are given and proven, and then the reason that to maximize the observe ability is to optimize the  signal detecting performances  is established.
			 Secondly, a dual relation between the  observability ellipsoid and the controllability ellipsoid, which volumes and radii are respectively with some inverse relations, is stated and proven. 			
			 Accordingly, the analytical computing equations for the volume of the two observability ellipsoids are got and some analytical shape factors of these ellipsoids are deconstructed. Based on these effective compting for the volumes, radii, and shape factors, analyzing and optimizing for the observe ability can be carried out. 
			  Thirdly, 
			 to compare rationally the state observe ability between the different systems or different system parameters, the normalization of the output variables, the state variables, and the system models are discussed. 
			 Finally, some numerical experiments and their results show the effectiveness of the computing and comparing methods for the observe ability. 
			
			
		 \end{abstract}
		
		
	 \end{frontmatter}
	
		
	 \section{Introduction} 
	 	
	Like the state controllability, the state observability, proposed in 1960's by R. Kalman, et al, is the one of the essential properties describing the dynamical systems and a footstone to support the development of the many fields, such as, control theorey and engineering, signal detection and processing, and so on \cite{Kailath1980}, \cite{Chen1998}. 
As we known, for the dynamical systems, the concepts can reveal deeply the possibility reconstructing the unmeasurable state variables by the measurable output variables, and also can be used to disclose the possibility recoving the state variables from the output signals polluting by noises. The concept and the corresponding analsysis method impels us to understand well the dynamical systems for designing the state detect systems and control systems.
	
	But, it is a great pity that the state observability is only a qualitative concept with two-value logic and the dynamical systems are distinguished as only two classes of systems, the observable systems and unobservability systems, according to the corresponding state observability criterion. The concept and criterion could not tell us the observe ability and the observe efficiency of the chosen output variables to estimate  the state variables, and the quantitative concept and analysis method on that are failed to be established. In fact, the quantitative concept and analysis method are very important for many practical fields, such as, control engineering, signal detecting and processing, and so on, and then many engineering problems are dying to the concepts and methods for getting the easier designing and implementing for signal detecting and processing systems, the better dynamcial process and closed-loop performance index for the control systems. For example, evaluating the observe ability and observe efficiency can help us to understand and solve the following important signal detecing and processing problems for the dynamical systems: 
	
	1) how to choose the signal detecting sensors and equipments (e.g., how much are the choosing sensor range and accuracy? how much is the A/D-converter's bit?), how to determine the output variables(e.g., choosing displacement variable, speed variable, or acceleration variable for some mechanics systems? choosing the electric current of the main circuit or excitation circuit for some DC motors), and how to place the location of the sensors in larger mechanical systems (e.g., mechanical cantilevers, bridges, side slopes, solar panels, etc) and the wireless sensor networks, for maximizing the observe ability and effiency for the engineering systems. 
	
	2) how to design and optimize the structure and technical parameters of the engineering plants to get the stronger observe ability and then to make designing and implementing easily the signal detecting and processing systems, control systems, and so on.
	
	3) how to determine the leaders, the sub-leaders, and the connections among the nodes in the networked control systems and formation system for maximizing the performance of these systems.

To summarize above, defining, quantifying and optimizing the observe ability are with the very greater signification for the signal detecting and processing, control theory and engineering, and so on.

In this paper, the definition on the observe ability for the state observe problem is studied systematically. It is worthing point out that the state observe problem and the state filtering/estimation problem are basically the same problems, and their differences are that one is free from noise and the another is not in their system model assumptions. Therefore, the observe problems discussed in this paper include the state filtering/estimation problems and the obtained results are e also applicable to the filtering/estimation problems. 
 Firstly, to define and compare rationally the state observe ability of the output variables between the different controlled plants or one controlled plant with the different system parameters, the normalization of the output variables, the state variables, and the system models are discussed. With the help of the normalization, the time-attribute observe ability with the unit output variables can be defined. And then, two theorems on the relations among the open-loop observe ability, the control strategy space (i.e., the solution space of the output variables for control problems), and the closed-loop time performance are purposed and proven. Based on that, the conclusion that it is necessary to optimize the observe ability for the practical engineering problems can be got. Finally, the simulation experiments show us the normalizing the variables and system models, and comparing the time-attribute contol ability between the different controlled palnts. 

 \section{Definition of The State Observe Ability}
 
 \subsection{What Is the State Observe Ability?}
	
	In this paper, the linear discrete-time (LDT) system $\Sigma(A,C)$ is as a sample for studying the definition and the analysis method on the state observe ability, and the obtained results can be generalized conveniently to other classes of dynamical systems. 
	
	In general, the LDT Systems can be formulated as follows:
 \begin{equation}
 \left \{
 \begin{array}{l}
x_{k+1}=Ax_{k}+Bu_{k} \\
y_{k}=Cx_{k}
 \end{array}
 \right. \quad x_{k} \in R^{n}, u_{k} \in R^{r}, y_{k} \in R^{m}
 \label{eq:a0201}
 \end{equation}
 \noindent where $x_{k}$, $u_{k}$ and $y_{k}$ are the state, input, and output
variables, respectively, and matrices $A \in R^{n \times n}$, $B \in
R^{n \times r}$ and $C \in
R^{m \times n}$ are the state matrix, input matrix and output matrix, respectively, 
in the system models. In fact, by the definition and criterion \cite{Kailath1980}, \cite{Chen1998}, the state observability, in theory, has no relation to the input variable $u_{k}$ and the input matrix $B$, and then the LDT models \eqref{eq:a0201} for studying the state observability can be represented as follows
 \begin{equation}
 \left \{
 \begin{array}{l}
x_{k+1}=Ax_{k} \\
y_{k}=Cx_{k}
 \end{array}
 \right. 
 \label{eq:a0201a}
 \end{equation}
 and can be abbreviated as $\Sigma(A,C)$. 
 
To investigate the state observability of the
LDT system \eqref{eq:a0201a}, the observability matrix and the observability Gramian matrix can be defined as follows
 \begin{align}
Q_{o,N} & = \left[ \begin{array}{c} C \\ CA \\ \vdots \\ CA^{N-1} \end{array} \right ]
 \label{eq:a0202} \\
G_{o,N} & = \sum _{i=0} ^{N-1} \left( CA^{i} \right)^T CA^{i}=Q_{o,N}^TQ_{o,N} \label{eq:a0203}
 \end{align}	
where $N \ge n$. That the ranks of the matrices $Q_{o,N}$ and $G_{o,N}$ are $n$, that is, the dimension of the state space the system \eqref{eq:a0202}, is the well-known sufficient and necessary criterion on the state observability for the LDT systems. 

Based on the definition equation \eqref{eq:a0203} of the observability Gramian matrix, the observability ellipsoid can be defined as follows \cite{DulPag2000} 
 \cite{Marx2004} \cite{Gress2005} \cite{Melykuti2007} \cite{Hinson2014} \cite{Birou2020}

 \begin{align}
E_{N} & = \left \{ z \left \vert z = G_{o,N}^{1/2} x_0 , \; \left \Vert x_0 \right \Vert_2 \le 1 \right. \right \} \notag \\
& = \left \{ z \left \vert z^T G_{o,N}^{-1} z \le 1, \; \forall z \in R^n \right. \right \} \label{eq:a0203a}
 \end{align}
It is worth noting that the variable $z$ in above definition equations is not a variable in the real physical systems and can be regarded as a virtual variable in a $n$-dimensional ($n$-D) virtual space. Hence, the observability ellipsoid is also a virtual geometry.

Based on these concepts and definitions, when the system $\Sigma(A,C)$ is state observable, we have the following relation between the initial state $x_0$ and the measured output sequence $ Y_N= \left \{ y_0, y_1, \dots , y_{N-1} \right \}$ without the noise polution
 \begin{align}
Y_{N} & = Q_{o,N}x_0 \label{eq:a0203d} \\
x_0 & = G_{o,N} ^{-1} Q_{o,N} ^T Y_N \label{eq:a0203e}
 \end{align}
and then, the theoretical observed value of the initial state $x_0$ can be computed by the measured output sequence $ Y_N$ as follows
 \begin{align}
 \hat x_0 & = G_{o,N} ^{-1} Q_{o,N} ^T Y_N \label{eq:a0203f}
 \end{align}
It is worth noting that the observing method for the initial state $x_0$ is not unique, and except the estimating equation \eqref{eq:a0203f}, there exist many observing methods for the state observe problems, for example, all kinds of the state observer designing methods. 

 When the output sequence $ Y_N$ is polluted by the noise sequence $ W_N= \left \{ w_0, w_1, \dots , w_{N-1} \right \}$ satisfying the following probability models
 \begin{align}
E \left(w_k \right)=0, \; \textnormal{and} \; E \left(w_k^T w_k \right)= \Lambda \label{eq:a0203g}
 \end{align}
where $ \Lambda$ is the covariance matrix of the noise $w_k$, the observed value $ \hat x_0 $ by Eq. \eqref{eq:a0203f} will satisfy the following convergence result
 \begin{align}
 \lim _{N \rightarrow \infty } E \left( \left \Vert \hat x_0 -x_0 \right \Vert _2 ^2 \right) \le \lim _{N \rightarrow \infty } 
 \left \Vert G_{o,N} ^{-1} \right \Vert _2 \left \Vert \Lambda \right \Vert _2 \label{eq:a0203h}
 \end{align}
Therefore, when
 \begin{align}
 \lim _{N \rightarrow \infty } 
 \left \Vert G_{o,N} ^{-1} \right \Vert _2 = 0 \label{eq:a0203h1}
 \end{align}
we have
 \begin{align}
 \lim _{N \rightarrow \infty } E \left( \left \Vert \hat x_0 -x_0 \right \Vert _2 \right) = 0 \label{eq:a0203h2}
 \end{align}
thaty is, the initial state $x_0$ can be estimated rightly without the observed error by Eq. \eqref{eq:a0203f}.

As we know, when one of the conditions
 \begin{align}
& \lim _{N \rightarrow \infty } \det \left( G_{o,N} \right) = \infty \label{eq:a0203h1a} \\
& \lim _{N \rightarrow \infty } \lambda _{ \textnormal {min} } \left (G_{o,N} \right ) = \infty \label{eq:a0203h1b} \\
& \lim _{N \rightarrow \infty } \textnormal {vol} \left( E_{N} \right) = \infty \label{eq:a0203h1c} \\
& \lim _{N \rightarrow \infty } r _{ \textnormal {min} } \left (E_{N} \right ) = \infty \label{eq:a0203h1d}
 \end{align}
is true, 
Eq. \eqref{eq:a0203h1} will hold, where $ \det \left( G_{o,N} \right) $ and $ \lambda _{ \textnormal {min} } \left (G_{o,N} \right )$ are the determinant value and the minimum eigenvalue of the nonnegative definite matrix $ G_{o,N} $, respectively; $ \textnormal {vol} \left( E_{N} \right) $ and $ r _{ \textnormal {min} } \left (E_{N} \right )$ are the volume and the smallest radius of the observability ellipsoid $E_N$, respectively.

Therefore, in papers \cite{VanCari1982}, \cite{Georges1995}, \cite{PasaZamEul2014}, and \cite{Ilkturk2015}, $ \det \left( G_{o,N} \right) $ and $ \lambda _{ \textnormal {min} } \left (G_{o,N} \right ) $, that is, the corresponding $ \textnormal {vol} \left( E_{N} \right) $ and $ r _{ \textnormal {min} } \left (E_{N} \right )$, can be used to quantify the observe ability of the output variables to the state space, and then be chosen as the objective function for optimizing and promoting the observe ability of the linear dynamical systems. But, it is very regret that the above quantitative studying on the state observe ability hasn't been made good progress for the following reasons.

1) It is lack of some good interpretation and definition about that. For example, 

i) which attribute observe ability and efficiency is described by the observability ellipsoid $E_N$? observe time optimal, observe accuracy optimal, or others?

ii) whether that the size and shape of the ellipsoid $E_N$, that is, $ \textnormal {vol} \left( E_{N} \right) $, $ r _{ \textnormal {min} } \left (E_{N} \right )$, and other factors, are bigger means that above quantitative observe ability is stronger or weaker, and then 
 means that the state observer is with the more accurate observed values, the faster observing processes, or the bigger state observing ranges owing to the limited ranges of the sensors?

2) It is lack of the analytical computing of the determinant $ \det \left( G_{o,N} \right) $ and eigenvalue $ \lambda _{ \textnormal {min}} \left (G_{o,N} \right ) $ (that is, the volume and the shortest radius observability ellipsoid $E_N$ ), these optimizing problems for the observe ability are solved very difficulty, and few achievements about that were made. 

In prectice, out of the need of the practical signal detecting and control engineering, quantifying and optimizing the observe ability are key problems in many engineering fields.

Next, the new decoding on the state observability and then a noval defintion on the state observe ability are discussed systematically as follows.

Considered the following LDT system with the observe noises in the output variables
 \begin{equation}
 \left \{
 \begin{array}{l}
x_{k+1}=Ax_{k} \\
y_{k}=Cx_{k}+w_k
 \end{array}
 \right. 
 \label{eq:a0201b}
 \end{equation}
where $w_k \in R^m$ is the observe noises in the output variables. Here, there is no restriction or probablity model assumption for these noises temporarily.

The so-called state observability is a property on the possiblity reconstructing the initial state $x_0$ from the measurable output variables $y_k$ with $w_k=0, \forall k \ge0$, that is, without the noise polution. Namely, the state observability is whether there exists a sole solution $x_0$ in the following equations with the measured output sequence $Y_N$.
 \begin{equation}
 \left \{
 \begin{array}{l}
y_0=Cx_0 \\
y_1=Cx_1=CAx_0 \\
 \dots \\
y_{N-1}=Cx_{N-1}=CA^{N-1}x_0
 \end{array}
 \right. 
 \label{eq:a0201c}
 \end{equation}
 The state observe ability proposed here is an ability that the more accurate estimation of the initial state $x_0$ from the measurable output variables $y_k$ with the noise polution. Namely, the state observe ability is a quantitative ability that the more accurate solution of the initial state $x_0$ in the following equations with the measured output sequence $ Y_{N}= \left \{ y_0, y_1, \dots , y_{N-1} \right \}, N \ge n$ poluting by the unmeasurable noise sequence $ W_N= \left \{ w_0, w_1, \dots , w_{N-1} \right \}$.
 \begin{equation}
 \left \{
 \begin{array}{l}
 y_0=Cx_0+w_0 \\
 y_1=CAx_0+w_1 \\
 \dots \\
 y_{N-1}=CA^{N-1}x_0+w_{N-1}
 \end{array}
 \right. 
 \label{eq:a0201d}
 \end{equation}
 
 In fact, the state observe ability proposed here is a quantification and promtion of the classical state observability based the observe equaution \eqref {eq:a0201d} of the output variables. To define propoerly and exactly the state observe ability of the output variables to the state variables, the noise distribution models will be discused firstly. 
 
 \subsection{The Noise Bounded Models}

 The classical noise model is described by the probability distributiuon and is strictly based on the sufficiently great sample numbers. For many engineering probelms, the sample number is limited and the exact probabilty distribution models are difficulty to be made. Latterly, out of the environment conditions and needs of the practical engineering problems, researchers in many fields, such as, statistics and control engineering, focus on the bounded probability models of the noises, and then the robust estimation and robust control with the less prior probability knowledge are established.
 
 Here, the distributions of the noise $w_k$ in the $k$-th sampling or the noise sequence $W_N$ can be chosen as the following bounded models.
 \begin{align}
 & \Omega_{a,p}(s): \; \Vert w_k \Vert _p \le s, \; \forall k \\
 & \Omega_{N,p}(s): \; \Vert W_N \Vert _p \le s 
 \end{align}
 where $s(s>0)$ is the bounded value; $p$ means the 1-norm, 2-norm, or $ \infty$-norm, corrsponding that the noise models are the strength bounded models, the energy bounded models, and the bounded models, respectively. In genernal, the value bounded value $s$ canbe normalizaed as 1, and then $ \Omega_{a,p}(s)$ and $ \Omega_{N,p}(s)$ can be noted as $ \Omega_{a,p}$ and $ \Omega_{N,p}$, respectively. 
 
 Furthermore, if the system $\Sigma(A,C)$ is a single-output system, the above bounded models can be rewritten as 
 \begin{align}
& \Omega_{a,p}(s): \; \vert w_k \vert \le s, \; \forall k,p=1,2,3 \\
& \Omega_{N,p}(s): \; \left ( \sum _{k=1} ^{N} \vert w_k \vert ^p \right ) ^{1/p} \le s, \;p=1,2 \\ 
& \Omega_{N, \infty}(s): \; \vert w_k \vert \le s, \; \forall k \
 \end{align}

In this paper, the observe ability based on the Gramian observability matrix and observability ellipsoid will be defined and discussed, and these studies will be with the relation only to the total energy bounded models $ \Omega_{N,2}(s)$. The observe abilities with other noise bounded models will be studied in next papers.

 \subsection{The feasible space for the state observe problems}
 
Based on the total energy bounded model $ \Omega _{N,2}(s)$, the so-called state observing problem for the initial state $x_0$ is to find a state observed value $ \hat x_0$ under constraint conditions
 \begin{align}
 \left \{ \hat w_k= y_k-CA^{k} \hat x_0, \; k= \overline{0,N-1} \right \} \in \Omega _{N,2}(s) \label{eq:b02015z} 
 \end{align} 
where $ \hat w_k$ is the residual output error in the output variable $y_k$ after state observed process.
Based on the above constraint conditions \eqref {eq:b02015z}, the feasible solution set of the state observed value $ \hat x_0$ can be defined as follows
 \begin{align}
S_2(N)= \left \{ \hat x_0 \left \vert \left \{ \hat w_k = y_k-CA^{k} \hat x_0, \; k= \overline{0,N-1} \right \} \in \Omega _{N,2}(s) \right. \right \} \label{eq:b02015} 
 \end{align} 
 In other words, the region of the residual error $ \hat w_k$ generated by the feasible solution set $S_2(N)$ of the state observed value $ \hat x_0$ are with same geometry region $ \Omega _{N,2}(s) $ as the noise $w_k$.

In fact, the goal of the state observing algorithms (or state observers/eatimators) is to find the optimal or satisfactory solution in the feasible solution sets $S_2(N)$, under some given error penalty functions about the output error $ \left \{ \hat w_k, \; k= \overline{0,N-1} \right \}$.

To discuss the accuracy of the observed value $ \hat x_0$, the observed error can be defines as follows
 \begin{align}
 \tilde{x}_0 & = \hat x _0- x _0 \label{eq:a0201e} 
 \end{align}
And then, by Eq \eqref{eq:a0201d} and the define equation \eqref{eq:b02015}, we have, 
 \begin{align}
 \hat w_k = CA^{k}x_0+w_k-CA^{k} \hat x_0=w_k- CA^{k} \tilde{x}_0 \label{eq:a0201e1} 
 \end{align}
Therefore, the feasible solution sets of the observed error $ \tilde{x}_0 $ can be written as
 \begin{align}
 S_2^e (N) & = S_2-x_0 \notag \\
& = \left \{ \tilde x_0 \left \vert \left \{ \hat w_k= w_k-CA^{k} \tilde x_0, \; k= \overline{0,N-1} \right \} \in \Omega _{N,2}(s) \right. \right \} \label {eq:b02016}
 \end{align}
 It is worth noting that the feasible solution set $S_2^e(N)$ is really existence but is unmeasurable and unknown because of the unknown initial state $x_0$ and unmeasurable noise $w_k$ for the practice observing problems. 
 
 In fact, because that the noise $w_k$ and the noise estimated value $ \hat w_k$ in the constraint condition \eqref {eq:b02015z} of the observing algorithms are with same geometry distribution $ \Omega _{N,2}(s)$,
 by Eq. \eqref {eq:b02016}, the feasible solution set $ S_2^e$ satisfies the following equation
 \begin{align}
 S_2^e (N) & = \left \{ \tilde x_0 \left \vert \left \{ CA^{k} \tilde x_0, \; k= \overline{0,N-1} \right \} \in \Omega_{N,2}(2s) \right. \right \} \label {eq:b02016a}
 \end{align}
 where $ \Omega _{N,2}(2s) = \Omega _{N,2}(s) \oplus \Omega _{N,2}(s) =2 \Omega _{N,2}(s)$. And then, by normalizing the bounded values of the noise models $ \Omega _{N,2}(2s) $, the feasible solution set $ S_2^e(N)$ can be defined as follows.
 
 \begin{defi} \label {de:a201} The normalizied space for the feasible solutions of the state observed error $ \tilde x_0$ according to the noise distribution model $ \Omega _{N,2}$ is defined by the following equations.
 \begin{align}
S_2^e(N) & = \left \{ \tilde x_0 \left \vert \left \{ CA^{k} \tilde x_0, \; k= \overline{0,N-1} \right \} \in \Omega _{N,2} \right. \right \} \notag  \\
& = \left \{ \tilde x_0 \left \vert \Vert Q_{o,N} \tilde x_0 \Vert ^2 _ 2 = \sum _{k=0} ^{N-1} \Vert CA^{k} \tilde x_0 \Vert_2 ^2 \le 1 \right. \right \} \label {eq:b02017z}
 \end{align}

 \end{defi}

 In fact, solving the observed value $ \hat{x}_0 $ by all state observing processes (or state observers/eatimators) is also finding a proper or optimal solution from the above feasible solution sets $ S_2 / S_2^e$ based on the given evaluation function on the output equation error $ \left \{ \hat w_k, \; k= \overline{0,N-1} \right \}$. Therefore, the small the size of the feasible solution sets $ S_2 / S_2^e$ for the same noise distribution model is, the more accurate the observed value $ \hat{x}_0 $ by all state observing processes maybe is, and then, we can say, the stronger the observe ability of the LDT system $\Sigma(A,C)$ is. Therefore, the sets $ S_2 / S_2^e$ can be regard as an index of the state observe ability and is used to define and quantify well the observe ability for the LDT system $\Sigma(A,C)$. Because that the feasible solution set $ S_2^e$
 of the observed error $ \tilde{x}_0 $ is independent from the initial state $x_0$ and is to be choosen more properly to define the state observe ability and then only $ S_2^e$ is analyzed and computed laterly.

 \subsubsection {The Geometry Shape of the Feasible Solution Space }

 For the feasible solution space $S_2^e(N)$ of the observed error $ \tilde{x}_0 $, similar to the controllability ellipsoid \cite{DulPag2000}, \cite{KURVARA2007}, \cite{PolNaKhl2008}, \cite{nkh2016}, \cite{cannkh2017}, \cite {zhaomw202003} and observability ellipsoid \cite {DulPag2000} \cite{Rahn2001} \cite{Marx2004} \cite{Gress2005} \cite{Melykuti2007} \cite{Hinson2014} \cite{Birou2020},  we have the following properties.
 
 \begin{property} \label{pr:a0701}
 	If the system $\Sigma(A,C)$ is the state observability, the feasible solution space $S_2^e(N)$ is a $n$-D ellipsoid. Otherwise, $S_2^e(N)$ is a region without the bounded in some $n-n_o$ linearly independent directions, where $n_o= \textnormal{rank} Q_{o,N}$.
 \end{property}
 
 \textbf{Proof.} (1) If the system $\Sigma(A,C)$ is not the state observability, the observability matrix $Q_{o,N}$ is not full rank for all $N \ge n$. Therefore, for some $n-n_o$ linearly independent vectors $f_i \in R^n$ with the any length $ \left \Vert f_i \right \Vert _ 2$, $i \in \overline{1,n-n_o}$,  we have,
 \begin{align} 
 & Q_{o,N}  f_i =
 \left[ \begin{array}{c}
 C \\
 CA \\
 \vdots \\
 CA^{N-1}
 \end{array} \right ] f_i
 =0 \\
 & \sum _{k=0} ^{N-1} \left \Vert CA^k f_i \right \Vert _2^2 =0 \le 1
 \end{align}
 And then, $f_i \in S_2^e(N) $. Because that the above $f_i$ in $S_2^e(N)$ is with the any length $ \left \Vert f_i \right \Vert _ 2$, $f_i$ can  approach to $\infty$. 
 
 Hence, If the system $\Sigma(A,C)$ is not the state observability, $S_2^e(N)$ is a region without the bounded in some $n-n_o$ linearly independent directions. Otherewise, $S_2^e(N)$ is a $n$-D finite geometry.
 
 (2) By the definition equation \eqref {eq:a0203}, we know, when the system $\Sigma(A,C)$ is the state observability, matrix $G _{o,N}$ is a positive define matrix and then the feasible solution space $S_2^e(N)$ is a $n$-D ellipsoid in $R^n$. 
 \qed

Therefore, by \textbf {Property \ref{pr:a0701}}, we have the following definition on the observability ellipsoid.
 
 \begin{defi} \label{defi:za01}
Fot the state observability system $\Sigma(A,C)$, the feasible space  $S_2^e(N)$ of the state observed error $\tilde {x} _0$ can be named as the state-observed-error observability ellipsoid, abbreviated as the observability ellipsoid.
\end{defi}

 As pointed above, the  observability ellipsoid $E(N)$ defined by \eqref {eq:a0203a} is only a virtual geometry in a virtual variable space, which doesn't existed in the physical systems and the real variable space. But, the observability ellipsoid $S_2^e(N)$  defined by the feasible space for the observe problems exist really and then the geometry $S_2^e(N)$ with the practical significance can reflet the state observability and observe ability better than $E(N)$.


  \subsection{The Definition of the State Observe Ability}
  
 
 From the point of view of geometry, the size of the observability ellipsoid $S_2^e(N)$ can be  described by its volume and some  shape factors, such as the radii, the farthest point, the nearest point, vertexes, outer sides, and so on. In fact, as analyzed above, the size of the observability ellipsoid $S_2^e(N)$, that is, the feasible solution space, is directly with  relation to the observe ability of the systems, and can be define and quantified that ability. Therefore, we have the following defintion on the state observe ability for the LDT systems.
 
 \begin{defi} \label {de:a1001}
 	The  observability ellipsoid $S_2^e(N)$ of the LDT system $ \Sigma \left( A, C \right)$ is defined as the state observe ability according to the normalizational noise distribution model $ \Omega_{N,2}$, characterised by its volume and shape factors. The smaller the volume and the shape of the observability ellipsoid $S_2^e(N)$ are, the smaller the feasible space of the observed error $ \tilde x_0$ is, and then the stronger the observe ability of the system is.
 \end{defi}
 
 Because that the observability ellipsoid $S_2^e(N)$ and its geometry size and shape are directly reflect the state observe ability of the LDT systems  under the noise distribution $ \Omega_{N,2}$, some geometry characteristics of $S_2^e(N))$, such as volume, radii, the farthest/nearest points, vertexes, outer sides, and so on, are need to be disucssed in detail and computed effectively. In this paper, the state observe ability based on the observability ellipsoid $S_2^e(N)$ under the noise distribution $ \Omega _{N,2}$ is studied firstly and other 2 observability regions under other noise distribution $ \Omega _{N,*}$ will be discussed in another papers.

 \subsubsection {The Robust Boundarys of the Observed/Estimated Errors}
 
 In fact, the boundary of the observability ellipsoid $S_2^e(N)$ is the maximum error of the state observe problem under the noise constraint $\Omega_{N,2}$, that is,  the robust boundary of the state observed error $\tilde{x}_0$. Therefor, the observability ellipsoid and its boundary are discussed in detail as follows.

 \begin{property} \label{pr:a0703}
	For the state observable system $\Sigma(A,C)$, the boundary of the observability ellipsoid $S_2^e(N)$ can be defined as follows
	 \begin{align}
	V(N) = \left \{ df 
	 \left \vert 
	d^2 \sum _{k=0} ^{N-1} \left \Vert CA^k f \right \Vert _2^2 =1, \; \forall f \in R^n \; \textnormal {and} \; \Vert f \Vert _2 =1 
	 \right. 
	 \right \} \label {eq:z0702b}
	 \end{align}
	
 \end{property}

 \textbf{Proof.} By the defintion equation \eqref {eq:b02017z}, we know, the intersection point $z=df$ between the unit direction $f$ and the boundary of the ellipsoid $S_2^e(N)$ must satisfy the following equation 
 \begin{align}
 \sum _{k=0} ^{N-1} \left \Vert CA^k df \right \Vert _2^2 = d^2 \sum _{k=0} ^{N-1} \left \Vert CA^k f \right \Vert _2^2 = 1 \label {eq:z0702c}
 \end{align}
where $d>0$ is the distance between the origin and the intersection point $z$, and $f$ is the unit-length vector as $ \Vert f \Vert _2 =1$. 
Therefore, the boundary of the ellipsoid $S_2^e(N)$ can be described by Eq. \eqref {eq:z0702b}.  \qed

The changes of the ellipsoid $S_2^e(N)$ and its robust boundary  with the incresing of the measured output $ \left \{y_k, k=0,1,2, \dots \right \}$ are also the key elements reflecting the onserve ability of the LDT systems.  By this, we have the following property for that.
 
 \begin{property} \label{pr:a0702}
	The observability ellipsoid $S_2^e(N)$ is a non increasing ellipsoid, that is,
 \begin{align}
S_2^e(j) \supseteq S_2^e(j+1) \label {eq:z0703}
 \end{align}	
Furthermore, if the system $\Sigma(A,C)$ is the state observability, we have,

1) $S_2^e(N)$ is a almost strictly reducing(decreasing) geometry, that is,
 \begin{align}
S_2^e(j) \supset S_2^e(j+N) \; \; \textnormal{and} \; \; \partial S_2^e(j) \cap \partial S_2^e(j+N) = \phi, \; \forall N \ge n \label {eq:z0704}
 \end{align}	
 
 2) $S_2^e(N)$ will converge to a $(n-m)$-D ellipsoid $S$ with the increasing of the sample number $N$, where $m$ is the number of eigenvalues with the positive real part for the matrix $A$.
 \end{property}

 \textbf{Proof.} (1) For any $z \in S_2^e(j+1)$, we have,
 \begin{align} \sum _{k=0} ^{j} \left \Vert CA^k z \right \Vert _2^2 \le 1
 \end{align}	
And then, for the same $z$, we have 
 \begin{align} \sum _{k=0} ^{j-1} \left \Vert CA^k z \right \Vert _2^2 \le 1
 \end{align}	
that is, $z \in S_2^e(j)$. Therefore, Eq. \eqref {eq:z0703} holds.

Furthmore, if the system $\Sigma(A,C)$ is the state observability, for any $j$ and $N \ge n$, we have
 \begin{align} 
 \textnormal {rank} 
 \left [ \begin{array}{c}
CA^j \\ CA^{j+1} \\ \vdots \\ CA^{k+N-1} 
 \end{array} \right] =n
 \end{align}	
And then, for any $z \ne 0$, we have 
 \begin{align} \sum _{k=j} ^{j+N-1} \left \Vert CA^k z \right \Vert _2^2 > 0
 \end{align}
Therefore, for any $z \in \partial S_2^e(j)$, we have,
 \begin{align} \sum _{k=0} ^{j-1} \left \Vert CA^k z \right \Vert _2^2 = 1 \; \; \textnormal {and} \; \;
 \sum _{k=0} ^{j+N-1} \left \Vert CA^k z \right \Vert _2^2 > 1
 \end{align}
that is, $z \notin S_2^e(j+N-1)$. Hence, Eq. \eqref {eq:z0704} holds.

(2) If the $m$ eigenvalues of the matrix $A$ are with the positive real part, the corresponding eigenvalues of the matrix $G_{o,N}$ will approach to $ \infty $, and then, by Eq. \eqref {eq:b02017z}, we know, the observability ellipsoid $S_2^e(N)$ will converge to the zero in the directions determine by the corresponding eigenvectors. Therefore, the observability ellipsoid will converge to a $n-m$-D ellipsoid in the subspace constructed by the other $n-m$ eigenvectors.
 \qed
 
\begin{figure}[htbp]
 	\centering
 	\includegraphics[width=0.7 \textwidth]{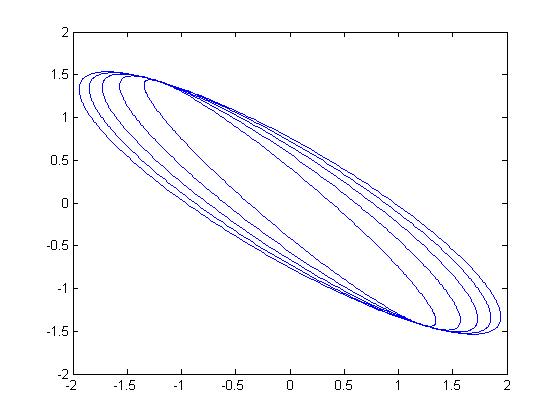} 
 	\caption[c]{The 2-D observability ellipsoid $S_2^e(N)$  when $N \in \{2, 6 \}$ \label {fig:zaa03}}	
 \end{figure}

  \textbf{Fig. \ref{fig:zaa03}} shows the 2-D  observability ellipsoid $S_2^e(N)$ generated by the matrix pair $(A, C)$ as follows
 $$
 A= \left[ \begin{array}{cc}
    0.900 &  -0.165 \\
0  &   0.350
 \end{array} \right], \; \; 
 C= \left[ \begin{array}{cc }
 1.00  & -1.30
 \end{array} \right]
 $$
 
 Because that the matrix pair $(A, C)$ is observabiity, the observability ellipsoid $S_2^e(N)$ in \textbf{Fig. \ref{fig:zaa03}} is strictly monotonic reduction
 with the increase of the sampling step $N$.
 
 In the next discussion, the system $\Sigma (A.C)$ are assumed always as a observable system and the geometry $S_2^e(N)$ is a $n$-D ellipsoid.

 \subsection{The Relation between of the Observe Ability and the Performance of The Detecting Systems }
 
 
 By above definitions and properties, we have some theorems and properties about the relation between the oberve ability and the performances of the signal detcting systems as follows.
 
 \begin{property}
 	 \label{pr:obsp0303}
 	If the LDT Systems \eqref{eq:a0202} with the noise constraint $ \Omega _{N,2}$ is state observable and 
 	the expecting robust boundary $e_b$ of the state estimated error $ \tilde x_0$, that is, the expecting maximum observed error, satisfy
 	 \begin{align} 
 	e_b \in S_2^e (N) \setminus S_2^e (N-1) \label{eq:a0227}
 	 \end{align}
 	the required minimum length of the measured output sequence $ \left \{y_0,y_1,y_2, \dots \right \}$ must equal to $N$, and but is not other smaller number, that is, the fewest observing sample number of the output variable $y_k$ is $N$.
 \end{property}


 \begin{theorem} \label{th:OBS0301} 
 	It is assumed that two LDT Systems $ \Sigma_1$ and $ \Sigma_2$ are state observable, and their observability ellipsoids are $S_{2}^{(1)} (N) $ and $S_2^{(2)} (N)$ respectively.
 	If we have
 	 \begin{align} \label{eq:a0228}
 	S_{2}^{(1)} (i) \subseteq S_{2}^{(2)} (i), \; \forall i \le N, 
 	 \end{align}
 	for the state observe problem and the same robust boundary $e_b \in S_{2}^{(1)} (N) \cap S_{2}^{(2)} (N) $ of the state observed error, the required minimum length of the measured output sequence for the system $ \Sigma_2$ is not more than that of $ \Sigma_1$, that is, there maybe exist some observe strategies with the less measured output sample number and the faster convergence speed of the observed error for the system $ \Sigma_2$.

 \end{theorem}

 \section {Computing on the Determinant of the Gramian Observability Matrix and the Volume of the Observability Ellipsoid}

By \textbf{Definition \ref {de:a1001}}, we know, the observe ability of the LDT systems can be described by the volume and some shape factors of the observability ellipsoid $S_2^e(N)$, such as, the maximum and minimum radii of the ellipsoid, and then the effective computing and analysis about these volume and shape factors
are the key problem for analyzing and optimizing the observe ability of the dynamical systems. Next, computing on that will be developed according to the two cases: 
finite-time case and infinite-time case.

 \subsection {Computing for the finite-time Observe ability Analysis }

By Eqs. \eqref {eq:a0203} and \eqref {eq:b02017z}, for the positive-define Gramian matrix  $G_{o,N}$ and the observability ellipsoid $S_2^e(N)$,   we have the following relations between their some features, such as, the ellipsoid volume and radii, the matrix determinant and eigenvalues.
 \begin{align}
r_i \left( S_2^e(N) \right) &= \mu_{i} ^{1/2} \left ( G_{o,N}^{-1} \right)= \mu_{n-i-1} ^{-1/2} \left ( G_{o,N} \right) , \; \; i=1,2, \dots , n 
 \label{eq:a2001} \\
 \textnormal{vol} \left( S_2^e(N) \right) & =H_n \prod _{i=1}^{n} r_{i} \left( S_2^e(N) \right) \notag \\
& =H_n \prod _{i=1}^{n} \mu_{i} ^{1/2} \left ( G_{o,N} ^{-1} \right) = H_n \det \left( G _{o,N}^{-1/2} \right) \notag \\
& = H_n  \prod _{i=1}^{n} \mu_{i} ^{-1/2} \left ( G_{o,N} \right) = H_n \left[ \det \left( G _{o,N}^{1/2} \right) \right ]^{-1} \label{eq:a2002}
 \end{align}
	where $r _{i} (S)$ and $ \mu _{i} (Q)$ are respectively the $i$-th maximum radius of the ellipsoid $S$ and $i$-the maximum eigenvalue of the matrix $Q$, and $H_n$ is the following volume-coefficient of the $n$-D hypersphere, 
 \begin{align} 
H_n & = \frac{ \pi^{n/2} } { \varGamma \left( \frac {n}{2}+1 \right)}
 \label{eq:a0218zz} 
 \end{align}	
where 
$ \varGamma (s)$ is the Gamma function defined as the follows.
 \begin{align} 
	 \varGamma(s)
	& = \left \{ \begin{array}{ll} 
	(s-1) \varGamma(s-1) & s>1 \\
	 \sqrt \pi & s=1/2 
	 \end{array} \right. \label{eq:a0219zz}
	 \end{align}	

Therefore, based on the computing of the Gramian observability matrix $G_{o,N}$ by Eq. \eqref {eq:a0203}, we can compute conveniently the volume and radii of the ellipsoid $S_2^e(N)$ by above Eqs. \eqref {eq:a2001} and \eqref {eq:a2002}, and then analyze and optimize the finite-time observe ability for the LDT systems.

 \subsection{The dual property between the observability ellipsoid and the reachability ellipsoid}

In papers  \cite{DulPag2000}, \cite{KURVARA2007}, \cite{PolNaKhl2008}, \cite{nkh2016}, \cite{cannkh2017},  \cite{zhaomw202003}, a controllability/reachability ellipsoid for the  LDT system 
$$ x_{k}=A_cx_{k}+B_cu_{k}$$
 with the total input energy bounded $ \left( \sum_{k=0} ^{N-1} \left \Vert u_k \right \Vert _2 ^2 \le 1 \right)$ is defined as follows
 \begin{align}
R(N) & = \left \{ x \left \vert x= \sum_{k=0} ^{N-1} A_c^kB_cu_k = Q_{c,N} U_N, \; \forall U_N: \Vert U_N \Vert _2 \le 1 \right. \right \} \label {eq:y0702} \\
& = \left \{ x \left \vert x G_{c,N} ^{-1}x \le 1 , \; \forall x \in R^n \right. \right \} \label {eq:y07021}
 \end{align}
where 
 \begin{align}
U_N & = \left[u_0^T,u_1^T, \dots,u_{N-1}^T \right]^T \\
Q_{c,N} & = \left[ B_c,A_cB_c, \dots,A^{N-1}_cB_c \right] \\
G_{c,N} & = \sum _{k=0}^{N-1} A_c^kB_cB_c^T\left (A_c^k \right) ^T=Q_{c,N} Q_{c,N}^T
 \end{align}

By Eq. \eqref {eq:y07021}, we know, the reachability ellipsoid $R(N)$ is also a $n$-D ellipsoid for the reachability system, and then its volume and radii can be computing by the following equations.
 \begin{align}
r_i \left( R(N) \right) & = \mu_{i} ^{1/2} \left( G _{c,N} \right) , \; \; i=1,2, \dots , n \label {eq:zz6003z1} \\
 \textnormal{vol} \left(R(N) \right) &=H_n \prod _{i=1}^{n} r_{i} \left( R(N) \right ) \notag \\
& =H_n \prod _{i=1}^{n} \mu_{i} ^{1/2} \left( G _{c,N} \right)= H_n \det \left( G _{c,N} ^{1/2} \right)
 \label {eq:zz6003z2}
 \end{align}

As we know \cite{Kailath1980} \cite{Chen1998}, the state observability of the LDT systems is equivalent to the state controllability/reachability of these dual systems. In fact, the observability ellipsoid $S_2^e(N)$ defined in this paper and the reachability ellipsoid $R(N)$   are also with the some form dual relations. For that, a dual property between two geometrys is stated and proven as follows.

 \begin{property} \label{pr:a0704} 
	If the observable system $\Sigma(A,C)$ and the reachable system $ \Sigma \left(A_c,B_c \right)$ are dual each other, that is, 
	 \begin{align} A_c=A^T, \; \; B_c=C^T \label{eq:aa0218z} 
	 \end{align}	
	the observability ellipsoid $S_2^e(N)$ and the rachability ellipsoid $R(N)$ generated respectively by the observability matrix $Q_{o,N}$ and the controllability matrix $Q_{c,N}$ are dual to each others, with the following dual relations about the volumes and radii of these elliposoid.
	 \begin{align}
	 \textnormal{vol} \left( S_2^e(N) \right) & = \frac {H^2_n} { \textnormal{vol} \left(R(N) \right)} \label {eq:z5001} \\ 
	r _{i} \left( S_2^e(N) \right) & = \frac {1} {r _{n-i-1} \left(R(N) \right)}, \; \; i=1,2,...,n \label {eq:z5002}
	 \end{align}
 \end{property}

It is worth noting that for the duality systems $ \Sigma \left(A, C \right)$ and $ \Sigma \left(A_c,B_c \right)$, the observability matrix $Q_{o,N}$ and the rachability ellipsoid $R(N)$ as two ellipsoids are duality each others and the duality relations are that the volume and radii of one ellipsoid are opposite to another ellipsoid, respectively.

By the above property, the computing and optimizing for one geomatry will be equivalent to that for the dual geometry. For example, computing and minimizing the volume and radii of observability ellipsoid $S_2^e(N)$ will be equivent to computing and maximizing the volume and radii of the dual geometry

 \textbf{Proof of Property \ref {pr:a0704}.} When the systems $\Sigma(A,C)$ and $ \Sigma \left(A_c,B_c \right)$ are duality each others, that is, Eq. \eqref {eq:aa0218z} holds, by the definitions of the Gramian observability matrix $G_{o,N}$ and the Gramian reachability matrix
$G_{c,N}$, we have following duality relations
$$ Q_{c,N}=Q_{o,N}^T, \; \; G_{c,N}=G_{o,N} $$
Therefore, we have 
 \begin{align}
 \mu_{i} \left( G_{o,N} \right) & = \mu_{i} \left( G_{c,N} \right ), \; \; i=1,2, \dots , n \\
 \det \left( G_{o,N} \right) & = \det \left( G_{c,N} \right ) 
 \end{align}
So, by Eqs. \eqref {eq:a2001}, \eqref {eq:a2002}, \eqref {eq:zz6003z1} and \eqref {eq:zz6003z2}, we know, 
Eqs. \eqref {eq:z5001} and \eqref {eq:z5002} hold. 
 \qed

 \subsection {Analytic Computing for the Infinite-time Observe Ability Analysis }

For the reachable  single-input LDT system $ \Sigma \left (A_c,B_c \right)$, when its all eigenvalues  are single root and the corresponding complex modulus are in the interval $[0,1)$, the analytic expressions of the determinant value of the Gramian controllabilty matrix $G_{c,N}$ and the volume of the reachability ellipsoid $R(N)$ for the system $ \Sigma \left (A_c,B_c \right)$ can be proven as the follows \cite{zhaomw202002} 
 \begin{align} 
 \det \left ( G_{c, \infty} \right ) &=
 \left \vert \det (P_c) \prod_{1 \leq i<j \leq n} \frac{ \lambda_{j}- \lambda_{i} }{ 1- \lambda_{i} \lambda_{j}} \right \vert ^2
 \prod_{i=1}^{n} \frac{ \left \vert q_iB_c \right \vert ^2}{1- \left \vert \lambda_{i} \right \vert ^{2}} \label{eq:a0216e} \\
 \textnormal {vol} \left (R ( \infty) \right) &= H_n \left [ \det \left ( G_{c, \infty} \right ) \right ] ^{1/2}
 \label{eq:a0217e}
 \end{align}
where matrix $P_c= \left[ p_1,p_2, \dots,p_n \right]$ is the diagonalization transformation matrix of the system matrix $A_c$; 
$ \lambda_{i}$, $p_i$ and $q_{i}$ are the $i$-th eigenvalue, and the corresponding unit right and left eigenvectors of matrix $A_c$, respectively.

Therefore, by \textbf{Property \ref {pr:a0704}} and the above analytical computing for the reachability ellipsoid, we have the following the theorem about the analytical volume expression of the observability ellipsoid $S_2^e( \infty)$ for the LDT system $\Sigma(A,C)$.

 \begin{theorem} \label {th:a3001}
For the observable single-output system $\Sigma(A,C)$, when its eigenvalues are single root and the corresponding complex modulus are in the interval $[0,1)$, the determinant of its Gramian observability matrix $G_{o,N}$ and the volume of its observability ellipsoid $S_2^e( \infty)$ can be expressed analytically and respectively as follows.
	 \begin{align} 
	 \det \left ( G_{o, \infty} \right ) &=
	 \left \vert \det (P) \prod_{1 \leq i<j \leq n} \frac{ \lambda_{j}- \lambda_{i} }{ 1- \lambda_{i} \lambda_{j}} \right \vert ^2
 \prod_{i=1}^{n} \frac{ \left \vert Cp_i \right \vert ^2}{1- \left \vert \lambda_{i} \right \vert ^{2}} \label{eq:a0216ez} \\
	 \textnormal {vol} \left ( S_2^e ( \infty) \right) &= H_n \left [ \det \left ( G_{o, \infty} \right ) \right ] ^{-1/2}
	 \label{eq:a0217ez}
	 \end{align}
	where matrix $P= \left[ p_1,p_2, \dots,p_n \right]$ is the diagonalization transformation matrix of the system matrix $A$; 
	$ \lambda_{i}$ and $p_i$ are the $i$-th eigenvalue and the corresponding unit right eigenvector of matrix $A$, respectively.
 \end{theorem}

By \textbf{Theorem \ref {th:a3001}}, we can compute analytically and conveniently the volume of the volume of the observability ellipsoid and analyze effectively the observe ability of the LDT system.

 \subsection{Decoding the observability ellipsoid and the observe ability}

 \subsubsection{The image observability ellipsoid}
 As pointed out in \textbf{Section 2}, the observability ellipsoid $E(N)$ define by Eq. \eqref {eq:a0203a} is not a real existence geometry in the state space of the physical systems and can be regraded as a virtual geometry. Next, the observability ellipsoid $E(N)$ will be redefined and extended as an image observability ellipsoid $S_2^i(N)$, a virtual geomety corresponding to the actual state-eatimated-error observability ellipsoid $S_2^e(N)$.
 
 \begin{defi} \label {de:a203}
 	For the state observability LDT system$\Sigma(A,C)$, the ellipsoid $S_2^i(N)$ defined by the following equations is an image geometry in a $n$-D virtual space correponding to the observability ellipsoid $S_2^e(N)$ and the state spece of the system, and can be called as an image observability ellipsoid. 
 	 \begin{align}
 	S_2^i(N)
 	& = \left \{ z \left \vert z^T G_{o,N}^{-1} z \le 1, \; \forall z \in R^n \right. \right \} \notag \\
 & = 	 \left \{ z \left \vert z = G_{o,N}^{1/2} x_0 , \; \left \Vert x_0 \right \Vert_2 \le 1 \right. \right \} \notag \\
 & = 	 \left \{ z \left \vert z = Q_{o,N}^T W_N , \; \left \Vert W_N \right \Vert_2 \le 1 \right. \right \} 
 	 \label{eq:a0203a1}
 	 \end{align}
where the variable $z$ is a virtual variable in a $n$-D space.
 	 \end{defi}
 
 By \textbf{ Definition \ref {de:a201} and \ref {de:a203}}, we know, the image observability ellipsoid $S_2^i(N)$ and the  observability ellipsoid $S_2^e(N)$, generated respectively by the positive define matrices $G_{o,N}$ and $G_{o,N}^{-1}$, are with some inverse relations, such as, these volumes, ellipsoid radii, and so on. The maximum radius of $S_2^i(N)$ is equal to the inverse of the minimum radius of $S_2^e(N)$, and meanwhile the minimum radius of $S_2^i(N)$ is equal to the inverse of the maximum radius of $S_2^e(N)$. In addition,
 the ellipsoid $S_2^i(N)$ as a geometry is fully equivalent to the reachability ellipsoid $R(N$). Therefore, the volume and the radii of the ellipsoid $S_2^i(N)$ can be computed analytically as follows
 \begin{align}
 r_i \left( S_2^i(N) \right) &= \mu_{i} ^{1/2} \left ( G_{o,N} \right), \; \; i=1,2, \dots , n 
 \label{eq:a2001z} \\
 \textnormal{vol} \left( S_2^i(N) \right) & =H_n \prod _{i=1}^{n} r_{i} \left( S_2^i(N) \right) \notag \\
 & =H_n \prod _{i=1}^{n} \mu_{i} ^{1/2} \left ( G_{o,N} \right) = H_n \det \left( G _{o,N}^{1/2} \right) \label{eq:a2002z}
 \end{align}	
And then,
when the eigenvalues of the single-output LDT system are single root and the corresponding complex modulus are in the interval $[0,1)$, the volume of its image observability ellipsoid $S_2^i( \infty)$ can be expressed analytically as follows.
 \begin{align} 
 \textnormal {vol} \left ( S_2^i ( \infty) \right) =	 \left \vert \det (P) \prod_{1 \leq i<j \leq n} \frac{ \lambda_{j}- \lambda_{i} }{ 1- \lambda_{i} \lambda_{j}} \right \vert
 \prod_{i=1}^{n} \frac{ \left \vert Cp_i \right \vert }{ \left (1- \left \vert \lambda_{i} \right \vert ^{2} \right )^{1/2}} \label{eq:a0216ez2}
 \end{align}
where the symbols $P$, $p_i$, and $ \lambda_{i}$ are same as in \textbf{Theorem \ref {th:a3001}}.

 \subsubsection{the shape factors of the observavility ellipsoid}

According to the volume computing equations  of the observability ellipsoid $S_2^e(N)$ and the image observability ellipsoid $S_2^i( \infty)$, some factors described the shape and size of theSE observability ellipsoidS, that is, the observe ability of the LDT systems, are deconstructed as follows.
 \begin{align} 
\left \{
\begin{array}{l}
F_1 = \left \vert \prod_{1 \leq j_{1}<j_{2} \leq n} \frac{ \lambda_{j_{2}}- \lambda_{j_{1}} }{ 1- \lambda_{j_{1}} \lambda_{j_{2}}} \right \vert  \\
F_{1,i,j} = \left \vert \frac{ \lambda_{i}- \lambda_{j} }{ 1- \lambda_{i} \lambda_{j}} \right \vert \\
F_{2,i} =  \left \vert C p_i \right \vert  \left ( 1- \left \vert \lambda_{i} \right \vert ^2 \right )^{-1/2} \\
F_{3,i}  = \left \vert C p_i \right \vert \\
r_i \left( S_2^e(N) \right) = r_{n-i+1} ^ {-1}\left ( S_2^i(N) \right) =\mu_{n-i+1} ^{-1/2} \left( G_{o,N} \right )
\end{array}
\right.
 \; \;i,j=1,2, \dots,n \label{eq:650}
 \end{align}
It is worth noting that excepting the radii $\left \{r_i \left( S_2^e(N) \right), i=\overline {1,n} \right \}$, above factors can be computed analytically. 

The above analytical factors $F_*$ can be called respectively as 
the shape(pole distribution factor) factor, the shape factor in 2-D section $x_i-x_j$, the side length of the circumscribed rhombohedral of the ellipsoid  $S_2^i( \infty)$ , and the modal controllability. In fact, the shape factor $F_1$ is also the eigenvalue evenness factor of the linear system, and can describe the oberve ability caused by the eigenvalue distribution. In addition, the modal observability factor $F_{3,i}$ have been put forth by papers \cite{chan1984} \cite{HamElad1988} \cite{ChoParLee2000} \cite{Chenliu2001}, and will not be discussed here.

 \subsection {The Shape Factors of the observability ellipsoids and the Eigenvalue Evenness Factor of the Linear System}

\begin{figure}[htbp]
	 \centering
	 \begin{minipage}[c]{0.45 \textwidth} 
		 \centering
		 \includegraphics[width=0.8 \textwidth ]{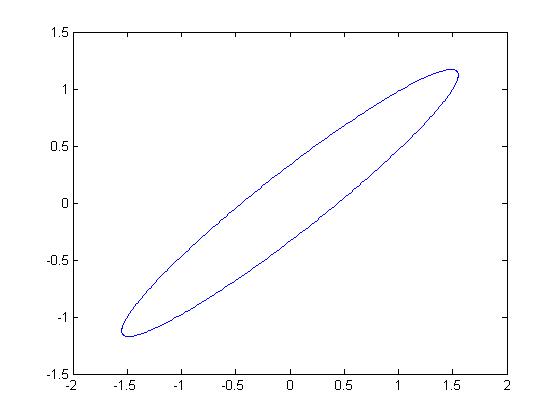} 
		 \\ (i) \footnotesize {the obser. ellip.}
	 \end{minipage}%
	 \begin{minipage}[c]{0.45 \textwidth}
		 \centering
		 \includegraphics[width=0.8 \textwidth]{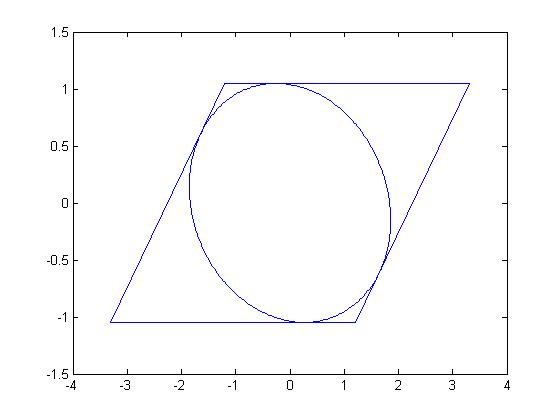}
		 \\ (ii) \footnotesize  {the image obser. ellip.}
	 \end{minipage}      \\   
 (a) \footnotesize {(0.3,0.9,0.8220)} \\ 
 	 \begin{minipage}[c]{0.45 \textwidth} 
 	\centering
 	\includegraphics[width=0.8 \textwidth ]{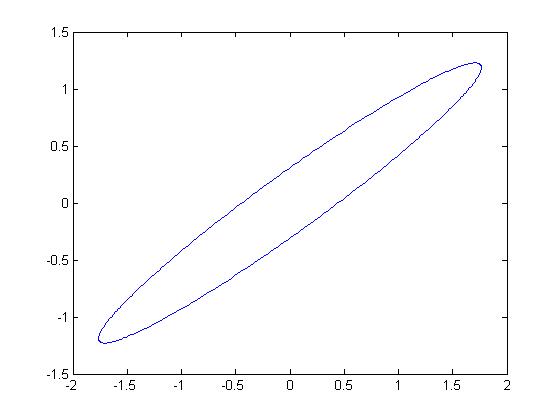} 
 	\\ (i) \footnotesize {the  obser. ellip.}
 \end{minipage}%
 \begin{minipage}[c]{0.45 \textwidth}
 	\centering
 	\includegraphics[width=0.8 \textwidth]{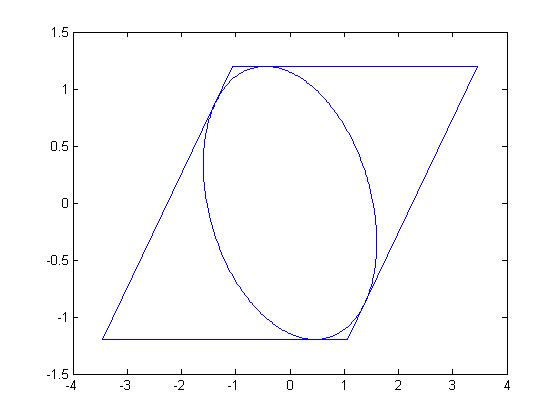}
 	\\ (ii) \footnotesize  {the image obser. ellip.}
 \end{minipage}      \\   
 (b) \footnotesize {(0.55,0.9,0.6931)} \\ 
 \begin{minipage}[c]{0.45 \textwidth} 
 	\centering
 	\includegraphics[width=0.8 \textwidth ]{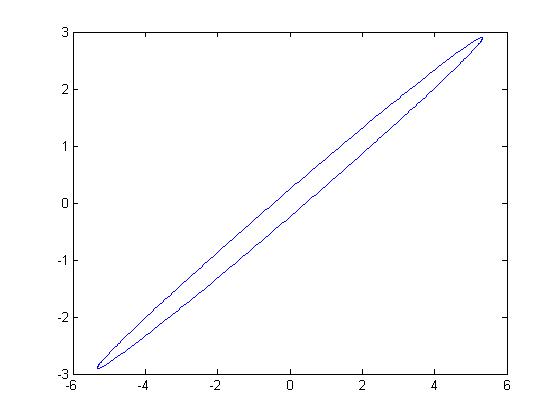} 
 	\\ (i) \footnotesize {the  obser. ellip.}
 \end{minipage}%
 \begin{minipage}[c]{0.45 \textwidth}
 	\centering
 	\includegraphics[width=0.8 \textwidth]{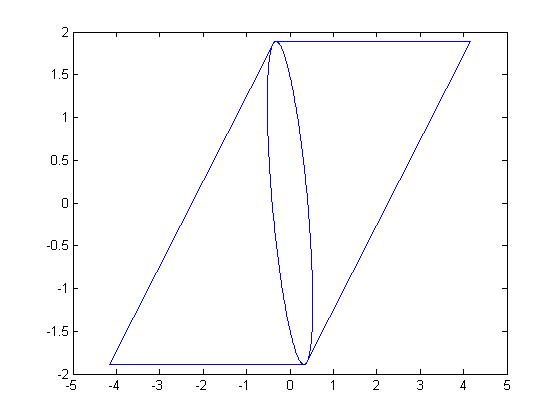}
 	\\ (ii) \footnotesize  {the image obser. ellip.}
 \end{minipage}      \\   
 (c) \footnotesize {(0.85,0.9,0.2128)} \\
	 \caption[c]{The 2-D observability ellipsoid with $( \lambda_1, \lambda_2,F_1)$ \label {fig:as01} }	
 \end{figure}

\begin{figure}[htbp]
	\centering
	\begin{minipage}[c]{0.45 \textwidth} 
		\centering
		\includegraphics[width=0.8 \textwidth ]{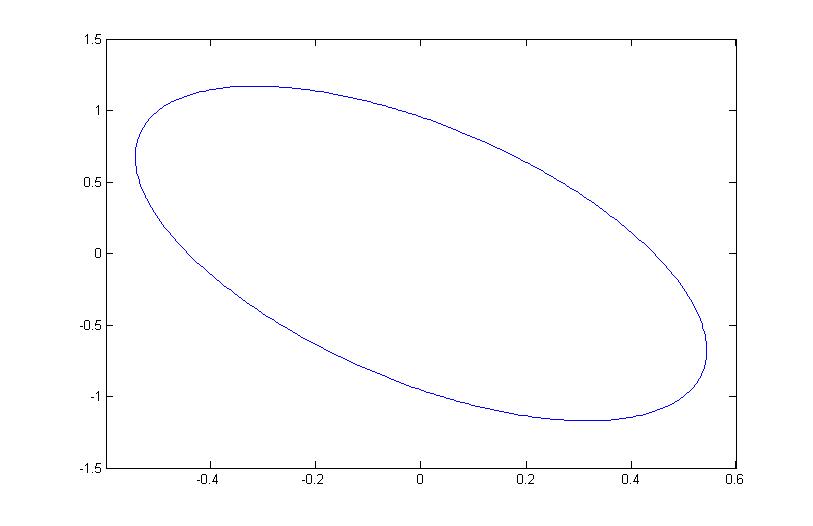} 
		\\ (i) \footnotesize {the  obser. ellip.}
	\end{minipage}%
	\begin{minipage}[c]{0.45 \textwidth}
		\centering
		\includegraphics[width=0.8 \textwidth]{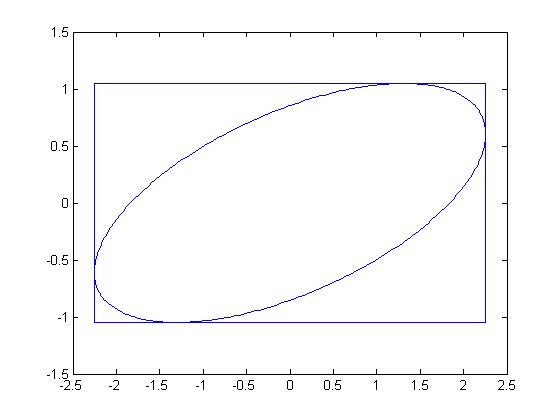}
		\\ (ii) \footnotesize  {the image obser. ellip.}
	\end{minipage}      \\   
	(a) \footnotesize {(0.3,0.9,0.8220)} \\ 
	\begin{minipage}[c]{0.45 \textwidth} 
		\centering
		\includegraphics[width=0.8 \textwidth ]{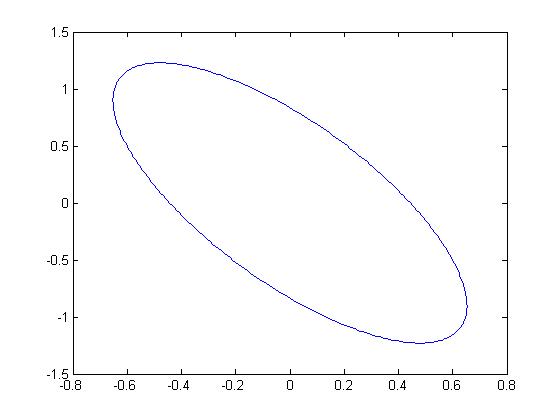} 
		\\ (i) \footnotesize {the  obser. ellip.}
	\end{minipage}%
	\begin{minipage}[c]{0.45 \textwidth}
		\centering
		\includegraphics[width=0.8 \textwidth]{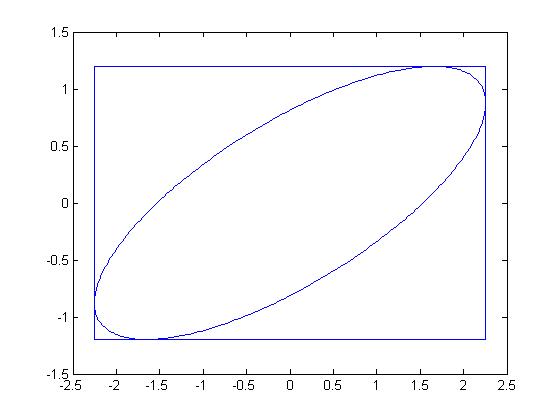}
		\\ (ii) \footnotesize  {the image obser. ellip.}
	\end{minipage}      \\   
	(b) \footnotesize {(0.55,0.9,0.6931)} \\ 
	\begin{minipage}[c]{0.45 \textwidth} 
		\centering
		\includegraphics[width=0.8 \textwidth ]{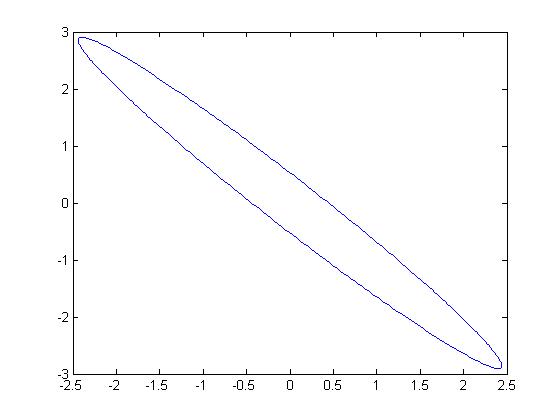} 
		\\ (i) \footnotesize {the  obser. ellip.}
	\end{minipage}%
	\begin{minipage}[c]{0.45 \textwidth}
		\centering
		\includegraphics[width=0.8 \textwidth]{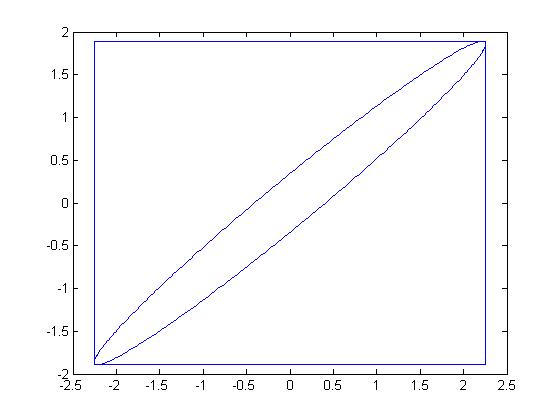}
		\\ (ii) \footnotesize  {the image obser. ellip.}
	\end{minipage}      \\   
	(c) \footnotesize {(0.85,0.9,0.2128)} \\
	\caption[c]{The 2-D observability ellipsoid in the eigen space with $( \lambda_1, \lambda_2,F_1)$ \label {fig:as02} }	
\end{figure}

Fig. \ref{fig:as01} shows the 2-D observability ellipsoid $S_2^e (16) $ and image observability ellipsoid $S_2^i (16) $, i.e., the sampling number $ N=16$, generated by the 3 matrix pairs $(A,c)$ that the matrix $A$ is with the different eigenvalues and matrix $c$ is a same vector, and Fig. \ref{fig:as02} shows the 2-D observability ellipsoids generated by the diagonal matrix pairs of these 3 matrix pairs $(A,c)$, that is, the  ellipsoids in Fig. \ref{fig:as02} are in the invariant eigen-space of the matrix $A$.

\begin{figure}[htbp]
	\centering
	\begin{minipage}[c]{0.45 \textwidth} 
		\centering
		\includegraphics[width=0.8 \textwidth ]{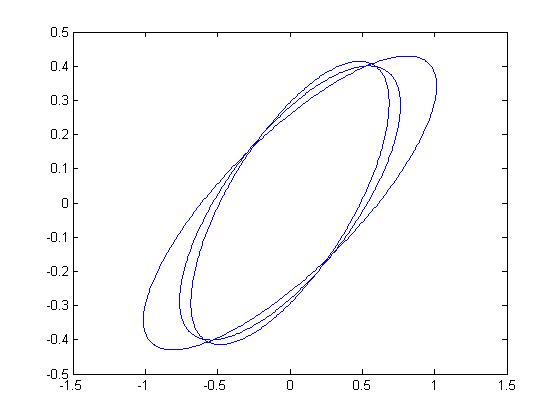} 
		\\ (a) \footnotesize {the  obser. ellip.}
	\end{minipage}%
	\begin{minipage}[c]{0.45 \textwidth}
		\centering
		\includegraphics[width=0.8 \textwidth]{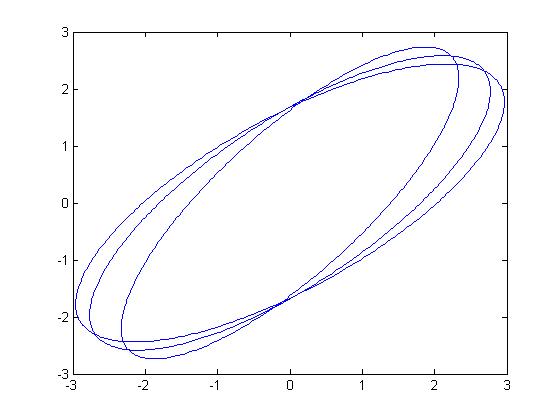}
		\\ (b) \footnotesize  {the image obser. ellip.}
	\end{minipage}      \\   
	\caption[c]{The 2-D observability ellipsoid with the comple eigenvalues  \label {fig:as03} }	
\end{figure}

Fig. \ref{fig:as03} shows the observability ellipsoids generated by the 3 matrices $A$ with the complex eigenvalues $0.9e^{\pm i \pi/24}$, $0.9e^{\pm i \pi/12}$ and $0.9e^{\pm i \pi/6}$ respectively, that is, the 3 pairs of eigenvalues  with same modulus but different complex phases. By the volume equation \eqref {eq:a0217ez} and Fig. \ref{fig:as03}, we see, when the complex phase is less $\pi/2$, the bigger the the complex phase 
is, the smaller the  observability ellipsoid is(meanwhile the bigger the image observability ellipsoid 
is), the smaller the observability ellipsoid is, and then the stronger the observe ability  of the LDT system is.

From these figures, we can see, when the two eigenvalues of the matrix $A$ are approximately equal, the minimum radius of the ellipsoid $S_2^e$ and $S_2^i$ will be approximately zero, and these  ellipsoids will be flattened. Similar cases are also for the $n$-D ellipsoid generated by the matrix pair.
Therefore, the distributions of all eigenvalues of the matrix $A$ are even, the ratio between the minimum and maximum radiie of the ellipsoids generated by the pair $(A,c)$ can be avoided as a small value and the ellipsoid will be avoided flattened, and then the observe ability can be enhanceed. 

The factor $F_1$ deconstructed from the volume computing equations \eqref{eq:a0217ez} and \eqref {eq:a0216ez2} can be used to describe the uniformity of the $n$ radii   of the ellipsoids $S_2^e$ and $S_2^i$ in $n$ eigenvectors. The bigger the value of the factor $F_1$, the bigger 
the ratios between the minimum and maximum radii of the ellipsoids $S_2^e$ and $S_2^i$ are, and then the bigger the volumes of the ellipsoids  $S_2^e$ and $S_2^i$ are.

Otherwise, the factor $F_1$ can be used to describe the evenness of the eigenvalue distribution of the LDT system $\Sigma(A,C)$. The bigger the value of the factor $F_1$, the smaller the observable ellipsoid $S_2^e$ of the system is (corrspondingly the bigger the image observable ellipsoid $S_2^i$ of the system is), and the stronger the observe ability of the LDT system is.

 \subsection {The circumscribed hypercube and circumscribed rhombohedral of the observability ellipsoid}

The factor $F_{2,i}$ is indeed the biggest distance of the image observability ellipsoid $ S_2^i$ in the $i$-D coordinate of the real-number eigen-space of the matrix $A$ (shown as in Fig. \ref{fig:as02} ), that is, the $n$ side lengths of the circumscribed hypercube of the ellipsoid $ S_2^i$ are $2F_{2,i},i= \overline{1,n}$. By the volume equation \eqref{eq:a0216ez2}, the volume of the ellipsoid $ S_2^i$ can be represented as the production of the volume $ \prod _{i=1} ^{n} F_{2,i}$ of the circumscribed hypercube and the shape factor $F_1$, and then the volume of the  observability ellipsoid $ S_2^e$ is the inverse of the above production.

Because that these expressions of the volume and the shape factors can describe accurately the size and shape of the ellipsoid generated by the matrix pair, i.e., the observe ability of the dynamical systems \cite{zhaomw202003}, these expressions can be used conveniently to be the objective function or the constrained conditions for the optimizing problems of the observe ability of dynamical systems.

 \section{Comparing the observe ability among the LDT systems}
 
To study rationally the observe ability of the output variables to the state variables in different dynamical systems, it is necessary to normalize the output variables, the state variables, and the system models, excepting for computing and analyzing their observability ellipsoids. Based on the normalization, the observe ability can defined and discussed in detail.

 \subsection{Normalization of the Variables and System Models}
	
In different practical signal detected or controlled plants, the physical dimensions, scales, value ranges of state variables and ths sensors for the output variables are different. Comparing rationally the observe ability of these different practical plants, or these different output variables/sensors, firstly, the output variables, the state variables, and then the system models must be normalized according to the practical signal detect or control problems. For example, to compare the observe ability between the two different systems or one system with some different parameters, the physical dimensions, scales, value ranges are necessary to be adjusted as a proper compatible values with some rationalness. Similarly, to compare the observe ability between the two different state variables, the dimensions, scales, ranges are also necessary to be adjusted as a proper compatible values. Next, two examples are discussed for showing these adjustment and normalization.

1) If only one output variable can be used to be placemented the sensor to detect signal for observing the state variables of a practical DC motor, which electric current variable, the main circuit or excitation circuit, is chosen as that for maximizing observe ability? The value ranges of these current variables and the ratios between the current variables and the speed variables of the motor must be adjusted to be with uniformity. Based on this, comparing with the different output variables is with rationalness and signification.

For example, if the rated values of the main circuit  $i_m(k)$ and excitation circuit  $i_e(k)$ as the output variable are respectively $i_m^*$ and $i_e^*$, that is, the output variables are respectively in the rated interval $[-i_m^*, i_m^*]$ and $[-i_e^*, i_e^*]$, to compare rational the observe ability of the output variables, the output variables of the system models for the two cases should be normalized respectively as 
 \begin{align} y_k=i_m(k)/i_m^* \; \textnormal{ and } \; y_k=i_e(k)/i_e^* \label{eq;a0304}
 \end{align}
where $y_k$ is the normalized output variable which the define domain is $[-1, 1]$. If the system models with the two output variables are respectively $ \Sigma \left( A_m, C_m \right)$ and $ \Sigma \left( A_e, C_e \right)$, the normalized system models are respectively as 
 \begin{align} \Sigma \left( A_m, i_m^*C_m \right) \; \; \textnormal {and} \; \; \Sigma \left( A_e, i_e^*C_e \right) \label{eq;a0305}
 \end{align}

 2) Which motor, DC motor with the excitation controller or AC motor with the variable frequency controller, can be determined to be used to the some electric speed control system for maximizing observe ability? 
for comparing and choosing the motors, the value ranges of their output variables, the state variables,  and the power of the electric energy  must be adjusted to be with uniformity. Based on this, comparing with the different plants is with rationalness and signification.
 
 For example, it is assumed that for two motors,  $i_q(k)$,  $v_q(k)$ and $ \dot v_q(k)$  are  the main circuit current,  the speed and acceleration, respectively, and meanwhile at the speed variables $v_q(k)$ of the motors are the output variables $y_q(k)$ , where 'q' is 'a' for AC motor or 'd' for DC motor.
 It is assumed that the rated values of these variables  are $\left(i_q^*, \; v_q^*, \; \dot v_q^* \right) $ and $\left( i_q, \; v_q, \; \dot v_q \right)$  are chosen as the state variables $x_q(k)$. 
 
 In fact, out of the practical engineering needs, the normalization of these variables and the system models for comparing the observe ability can be divided into  the following two cases: all variables are expected to work in their own rated intervals or the same intervals.
 
 i) For comparing the observe ability of two motors when the all variables in their rated intervals, the output variables, the state variables, and  the system models $ \Sigma \left( A_q, B_q \right)$ should be normalized as 
 \begin{align} 
 y_q(k)/y_q^*, \;\; P_q^{-1}x_q(k), \;\; \textnormal {and} \;\; \Sigma \left( P_q^{-1} A_qP_q, y_q^*C_qP_q \right)  \label{eq:a0207}
 \end{align}
 where the normalization matrix $P_q= \textnormal {diag} \left \{i_q^*, v_q^*, \dot v_q^* \right \}$. And then, the all output and state variables are in the define domain $[-1, 1]$.
 
 ii) For comparing the observe ability of two motors which the all  variables are in the same expecting interval as follows
 $$[-i_s, i_s], \;[-v_s, v_s], \;[- \dot v_s, \dot v_s], \;$$
 the variables and the system models should be normalized as 
 \begin{align} 
 y_q(k)/y_q^*, \;\; P_s^{-1}x_q(k), \;\; \textnormal {and}
 \;\; \Sigma \left( P_s^{-1} A_qP_s, y_q^*C_qP_s \right) \label{eq:a0210}
 \end{align}
 where the normalization matrix 
 $
 P_s= \textnormal {diag} \left \{i_s^*, v_s^*, \dot v_s^* \right \} 
 $.
 
 After the normalization of the system models as above, analyzing, comparing and optimizing the observe ability between the different dynamical systems are with rationality. The above normliztion methods are also applied to other signal detected systems and controlled plants. 
 
 \subsection { The Procedure for Compring the Observe Ability}
 
 Acoording the above analysis and discussion, we have the following procedure for compring the observe ability between the two different systems or one system with some different parameters.
 
  \begin{procedure} \label {proc:a001}  The Procedure for Compring the Observe Ability:
  	
  	Step 1. Normalizing the output variables, state variables, and system Models
  	
  	Step 2. Computing the volume, radii, and shape factors of the  observability ellipsoid $S_2^e(N)$ (or the image observability ellipsoid $S_2^i(N)$ )
  	
  	step 3. Computing the observe ability  based on one of the following methods
  	
  	1) Based on the Lower bound constraint on the radii and shape factors, let the observe ability be the volume
  	
  	2) Let  the observe ability be some weighting sum of the volume,  radii, and shape factors.
  	
  	Step 4. Based on the computing results of the observe ability, choosing the output variabls or the system parameters of the signal detecting systems.
  	
  \end{procedure}

 \section {Numerical Experiments}(temporarily unavailable)

 \section {Conclusions}

		In this article, the definition on the observe ability and its relation to the signal detecting  performance  are studied systematically for the linear discrete-time(LDT) systems. Firstly, to define and  analyze the observe ability for the practical systems with the  measured noise, six kinds of bounded noise models are classified. For the noise energy bounded case, the  observability ellipsoid and the image 
	observability ellipsoid based on the Gramian observability matrix are defined and then a novel concept on the LDT systems, called as the observe ability, is defined. Based on that, some theorems and properties about the reletion between the observability ellipsoid (observe ability) and the signal detecting performances  are given and proven, and then the reason that to maximize the observe ability is to optimize the  signal detecting performances  is established.
	Secondly, a dual relation between the observability ellipsoid and the controllability ellipsoid, which volumes and radii are respectively with some inverse relations, is stated and proven. 			
	Accordingly, the analytical computing equations for the volume of two observability ellipsoids are got and some analytical shape factors of these ellipsoids are deconstructed from the analytical volume equations. Based on these effective compting for the volumes, radii, and shape factors, analyzing and optimizing for the observe ability can be carried out. 
	Thirdly, 
	to compare rationally the state observe ability between the different systems or one system with the different system parameters, the normalization of the output variables, the state variables, and the system models are discussed. 
	Finally, some numerical experiments and their results show the effectiveness of the computing and comparing methods for the observe ability.

	 \bibliographystyle{model1b-num-names}


 \end{document}